\documentclass[a4paper,10pt,twoside]{cpc-hepnp}

\usepackage{multicol}
\usepackage{graphicx}
\usepackage{booktabs}
\usepackage{amssymb,bm,mathrsfs,bbm,amscd}
\usepackage[tbtags]{amsmath}
\usepackage{lastpage}
\usepackage{enumerate}

\begin{document}

\fancyhead[c]{Submitted to Chinese Physics C} \fancyfoot[C]{\small \thepage}


\title{Development of large-area quadrant silicon detector for charged particles
\thanks{Supported by National Basic Research Program of China (Grant No.2013CB
834404) and National Natural Science Foundation of China (Grant No. 10727505,
10735100, 11375268)}}

\author{%
      P.~F.~Bao$^{1}$
\quad C.~J.~Lin$^{1;1)}$\email{cjlin@ciae.ac.cn}%
\quad F.~Yang$^{1}$
\quad Z.~Q.~Guo$^{2}$
\quad T.~S.~Guo$^{2}$\\
\quad L.~Yang$^{1}$
\quad L.~J.~Sun$^{1}$
\quad H.~M.~Jia$^{1}$
\quad X.~X.~Xu$^{1}$
\quad N.~R.~Ma$^{1}$\\
\quad H.~Q.~Zhang$^{1}$
\quad Z.~H.~Liu$^{1}$
}
\maketitle

\address{%
$^1$China Institute of Atomic Energy, Beijing 102413, China\\
$^2$Beijing Kelixing Photoelectric Technology Co., Ltd, Beijing 102413, China\\
}

\begin{abstract}
  The quadrant silicon detector, a kind of passivated implanted planar silicon
detector with quadrant structure on the junction side, gained its wide application
in charged particle detection. In this paper, the manufacturing procedure,
performance test and  results of the quadrant silicon detector developed recently
at the China Institute of Atomic Energy are presented. The detector is about
300 $\mu$m thick with a 48$\times$48 mm$^{2}$ active area. The leakage current
under full depletion bias voltage of -16 V is about 2.5 nA, and the raising time
is better than 160 ns. The energy resolution for 5.157 MeV $\alpha$-particle is
around the level of 1\%. Charge sharing effects between the neighboring quads,
leading to complicated correlations between two quads, were observed when
$\alpha$ particles illuminated on the junction side. It is explained as a result
of distortion of electric field of inter-quad region. Such events is only about
0.6\% of all events and can be neglected in an actual application.
\end{abstract}

\begin{keyword}
Quadrant silicon detector,\quad Passivated implanted planar silicon,\quad
Energy resolution,\quad Charge sharing effect
\end{keyword}

\begin{pacs}
07.77.-n,\quad 29.40.Wk
\end{pacs}

\footnotetext[0]{\hspace*{-3mm}\raisebox{0.3ex}{$\scriptstyle\copyright$}2013
Chinese Physical Society and the Institute of High Energy Physics
of the Chinese Academy of Sciences and the Institute
of Modern Physics of the Chinese Academy of Sciences and IOP Publishing Ltd}%

\begin{multicols}{2}

\section{Introduction}
  Silicon detectors have been widely used in nuclear physics and high energy
physics during the past decades thanks to the improvement of modern semiconductor
technology~\cite{knoll,lutz,leda,walton}. With their good energy resolution,
fast time response and reliable stability, silicon detectors have been applied
to energy spectrum measurement, time signal pickup, and particle identification
as well as tracking systems, such as the use in the complete-kinematics
measurement of two-proton emission from extremely proton-rich
nuclei~\cite{lin2009,xu2010,xuplb}, Silicon Ball for the research on the weakly
bound nuclei close to particle drip line~\cite{siball}, and
MUST/MUST2~\cite{must,must2} for radioactive beam experiments. Among them,
Passivated Implanted Planar Silicon (PIPS) detectors fabricated in the
``Planar'' process~\cite{kemmer80,kemmer84} have enjoyed widespread adoption.

  PIPS detectors, in most applications, have some advantages over traditional
Silicon Surface Barrier (SSB) detectors~\cite{knoll,lutz}. All the junction
edges of the PIPS detectors are buried in the process of oxide passivation,
which can achieve better stability and lower reverse leakage current, while
the SSB detectors have crude junction edges. In addition, junctions formed
by well controlled ion-implantation provide thin entrance windows and can
also reduce the leakage current. Low leakage current makes noise reduction.
Moreover, thin entrance window means thin dead layer, which brings not only
improvement of energy resolution but also reduction of energy straggling.
Therefore, it enables the possibility of closer detector-source distance
to achieve compact geometry and high efficiency~\cite{bergmann}.

  Despite of the excellence of their response, however, PIPS detectors have
shortcomings as well, especially for ones of large area. Large capacitor resulted
from large area will in turn lead to slow rise time and small voltage output
with equal quantity of charge ionized. Besides, detector capacitor is also one
of the non-negligible noise causing factors. In the applications that demand time
signal pick-up and high signal-to-noise ratio, one can use a quadrant PIPS to
replace large-area one in view of that the area is reduced to a quarter.
Quadrant silicon detectors (QSDs) also have other advantages, for instance,
a certain position resolution and the reduction of the count rate of readout
electronics. There are already commercially available quadrant silicon
detectors~\cite{micron} and they have been used in ISOLDE Silicon Ball~\cite{siball}.

  The QSD has been successfully developed at the China Institute of Atomic
Energy (CIAE). The manufacturing procedure, test procedure, and electronics
as well as detection performance are presented in detail. Afterwards,
the complicated correlations between two neighbouring quads, as a result
of charge sharing effects, are discussed.

\section{Manufacturing procedure}
From the manufacturing point of view, the planar process was developed for
producing Integrated Circuits (IC), and it has also been applied to the fabrication
semiconductor detectors nowadays~\cite{knoll,kemmer84}. The process is
complicated, that combines many crucial techniques such as oxide passivation,
photolithography and ion implantation. The whole production flow of our
PIPS-QSD goes as follows:

\begin{enumerate}[a)]
 \item Polish and clean. The fabrication of detectors start from a 4 inches,
 300 $\mu$m thick, high-resistivity (greater than 10 k$\Omega \cdot$cm),
 high-purity, n-type silicon wafer. The wafer should be polished and cleaned
 at the beginning.
 \item Oxide passivation. The polished Si wafers are passivated by thermal
 oxidation at temperature up to 1030$^{\circ}$C. O$_2$ gas stream flow past the
 Si wafers and then about 600 nm thick oxide layer (SiO$_2$) come into
 being on the wafer surface.
 \item Photolithography. Four entrance windows, corresponding to the four
 quadrants of the finished detector, should be opened. Each one has the
 area of 24$\times$24 mm$^2$. The oxide layers on those areas are removed
 with the photolithographic equipment and  proper etching techniques.
 \item Ion implantation. The junction side is formed by boron ions
 implantation within the window, while the ohmic side is implanted with
 phosphorus ions. Therefore, a thin layer of pure silicon on surface are
 doped into p-type on junction side and n-type on ohmic side, respectively.
 Proper energy and dose of boron ions are typically 15 keV and 5$\times 10^{14}$
 ions/cm$^2$.
 \item Annealing. Thermal annealing is one of the most effective methods to
 remove the radiation damage in the implanted layers in the premise of
 guarantee low leakage current of detectors. The implanted Si wafers should
 be annealed in dry N$_2$ gas at 600$^{\circ}$C.
 \item Al metallization. Both the front and rear surfaces should be evaporated
 by aluminum for ohmic electrical contacts. The aluminum layer should be very
 thin, for the energy straggling of incoming particles in aluminum layer
 contributes to the noise.
 \item Al patterning at the front. The aluminum on the oxide layers should be
 removed using photolithographic technique.
 \item Alloy. The aluminum electrode should be alloyed at 380$^{\circ}$C.
 \item Encapsulation. The last step is separating each individual detectors and
 then encapsulating. The detector chip is mounted on the print circuit board (PCB)
 and the golden wires connect the aluminum electrodes with readout contacts on
 PCBs using ultrasonic soldering.
\end{enumerate}

  The profile view of the QSD is depicted in Fig.~\ref{fig1}. Active area of
the detector is 48$\times$48 mm$^2$, which is divided into four quads
(24$\times$24 mm$^2$) by the SiO$_2$ bar of 0.1 mm wide.

\begin{center}
   \includegraphics[bb=0 0 630 310, width=7cm]{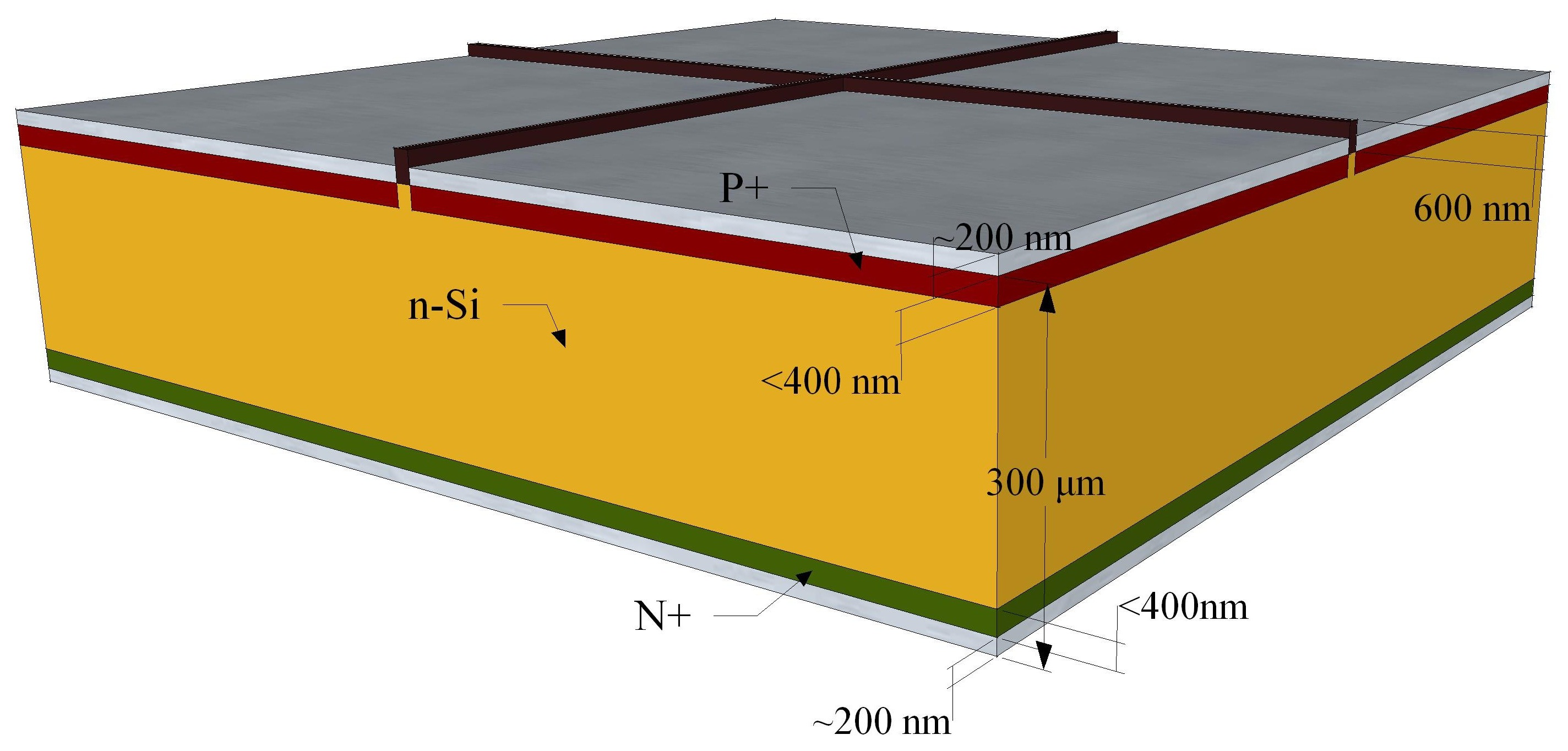}
   \figcaption{\label{fig1}The profile view of the QSD.}
\end{center}

\section{Electronics performance}
\subsection{Electronic setup}
  The test was performed with $^{241}$Am and $^{239}$Pu $\alpha$ sources.
As shown in Fig.~\ref{fig2}, the detector and the preamplifiers were put in a
vacuum chamber. $\alpha$-particles from radioactive source irradiated the QSD,
then the produced charges were collected by the charge sensitive
preamplifiers. With the aim of multi-channel integration, low noise and high
signal-to-noise ratio, the preamplifiers were well designed on a PCB to direct
connection with detectors. Performance of the preamplifier is excellent, with
bandwidth around 245 MHz and noise better than 5 nV/$\sqrt{Hz}$. Negative bias
voltage was applied to the junction side through 100 M$\Omega$ resistor built
in the AC-coupled preamplifier while the ohmic side was directly earthed.
The injected charges were integrated on the feedback capacitor into a current
pulse, which was then shaped by a spectroscopy amplifier CAEN N1568A. Timing
signals were produced by the build-in constant fraction discriminator (CFD) of
the amplifier, as a trigger to generate the gate signal for V785 ADC. Based on the
VME data acquisition system, V785 ADC converted the height of pulse into a digital
signal to save on a PC for further data analysing. The sequence of the number
of four quads is also shown in Fig.~\ref{fig2}. Hereinafter, we call
Quad.1 as Q1 for convenience, and the same for the other ones.

\begin{center}
\includegraphics[bb=0 0 629 352,width=7.5cm]{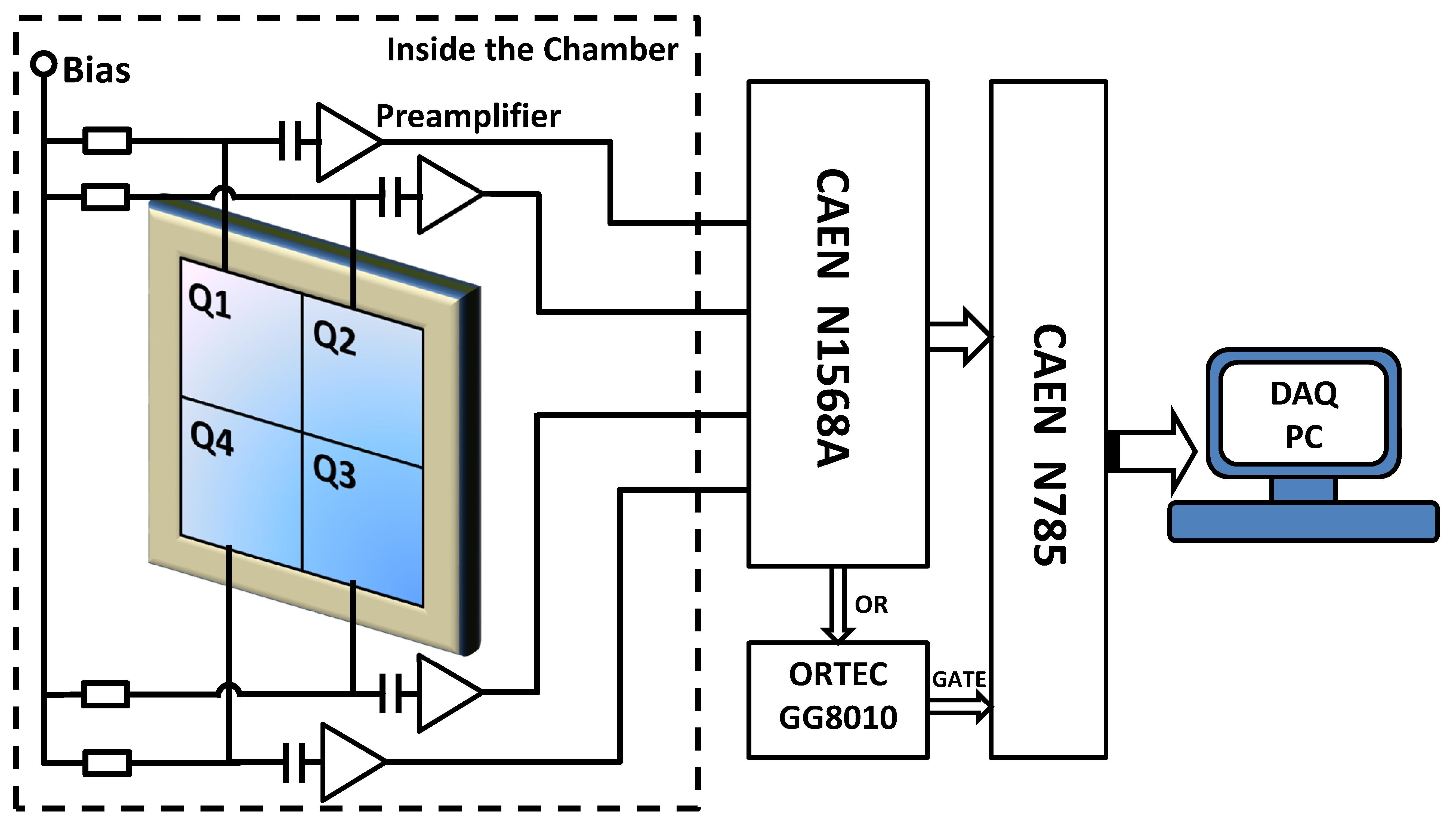}
\figcaption{\label{fig2}The block diagram of electronic setup.}
\end{center}

\subsection{Leakage current}
  The bias voltage applied to the detector is supplied by a Keithley 2602A
Source Meter, total leakage currents of four quads can be measured simultaneously.
Fig.~\ref{fig3} presents the curve of leakage currents of the QSD as a
function of the bias voltage. The test was performed at room temperature
25 $^{\circ}$C. When the detector is fully depleted at the bias voltage -16 V,
the leakage current of each quad is as low as 2.53 nA. Even under the bias
of -20 V, the leakage of each quad is merely 5.01 nA. During the whole
test procedure, the leakage currents are roughly constant, showing a good
stability.

\begin{center}
\includegraphics[width=7cm]{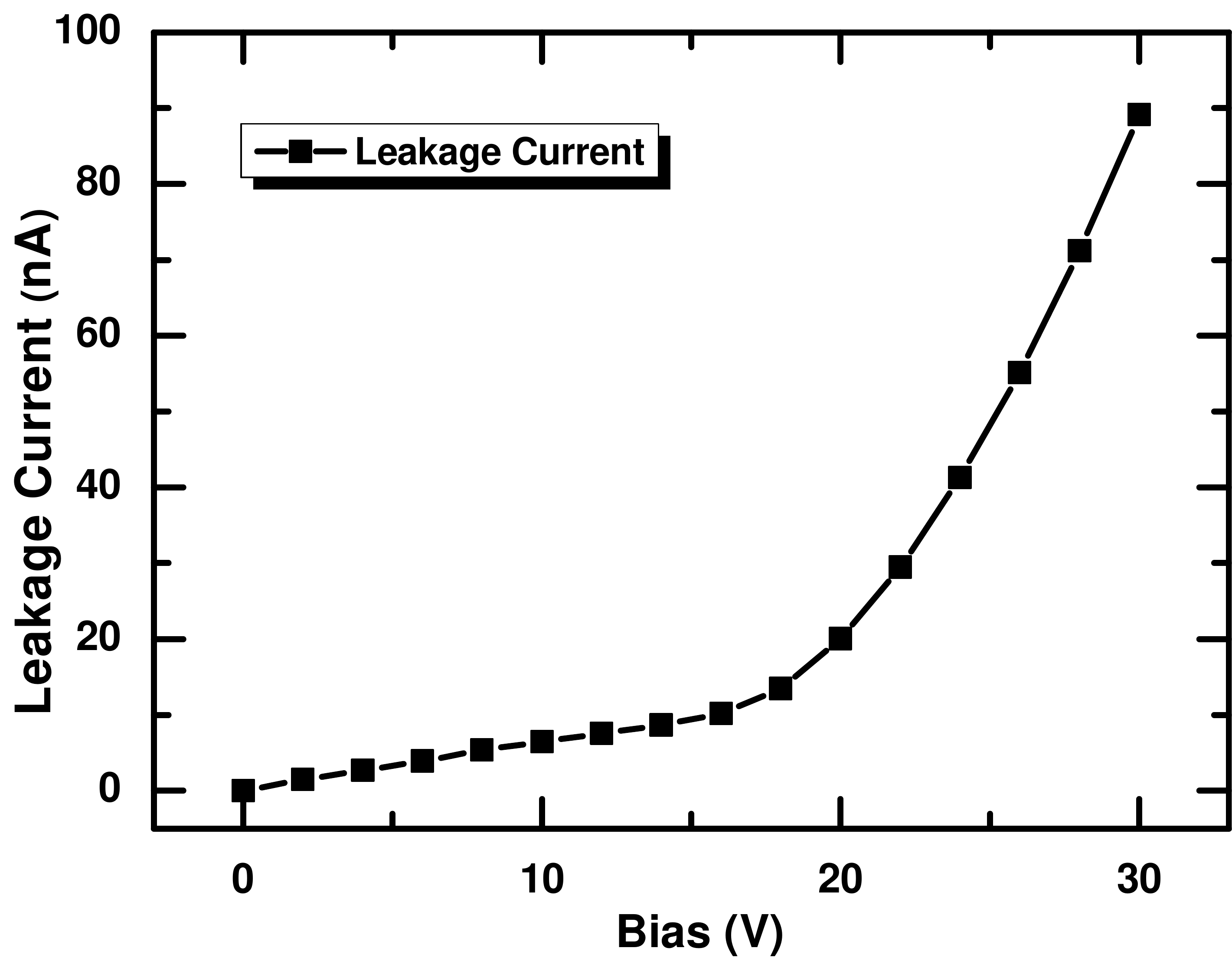}
\figcaption{\label{fig3}I-V curve of the QSD. The leakage currents are plotted as
a function of bias voltage. The leakage currents here are the sum of currents from
four quads.}
\end{center}

  Leakage current is a significant factor that should be considered for
operating almost all the semiconductor junction detectors. For one thing,
great magnitude of  leakage currents could bury weak signals to be measured.
For another, fluctuations of the leakage currents contribute to the
electronic noise, which broaden the overall peak and consequently reduce
the energy resolution. It is known that leakage currents originate from the
surface and bulk volume of the detector. The bulk leakage results from
the diffusion of the minority carriers and the thermal generation of
electron-hole pairs, both of which show little connection with the
manufacturing process. Different from this, surface leakage currents,
related to the large voltage gradient at the junction edges, are dependent
on the encapsulation of detector and the contamination on the detector
surface. From this point of view, the low level of the leakage currents
of these QSDs represents the fabricating process technic level of the
detector to some extent.

\subsection{Energy and timing responses}
  The output energy signals of preamplifiers generated by $\alpha$-particles
illuminating onto the detector are injected into an oscilloscope, thus the
pulse amplitude and rise time measurements are accessible. The upper panel
in Fig.~\ref{fig4} shows the pulse amplitude and rise time as a function
of reverse bias voltage when $\alpha$ particles irradiating on the junction
side of the detector. The bias increases with a step of 2 V from 0 V to -30 V.
As shown in Fig.~\ref{fig4}, the pulse goes high and rises fast while the bias
goes up from 0 V to -10 V. Once the bias is increased over -12 V, the pulse
height is roughly constant and remains 89 mV while the rise time decreases
slowly and levels off at higher voltage.

\begin{center}
\includegraphics[width=7cm]{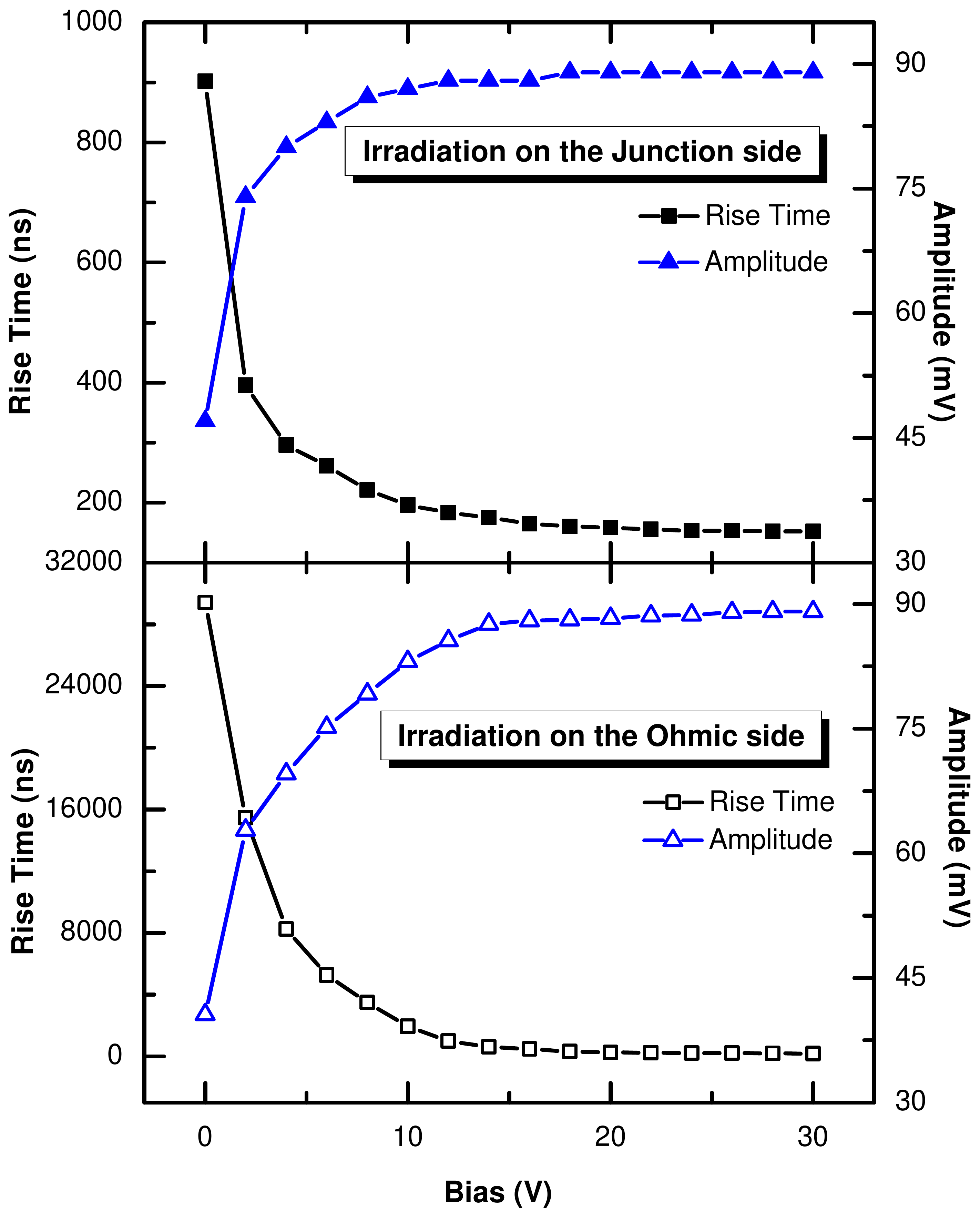}
\figcaption{\label{fig4}The upper panel shows the pulse amplitude and rise time as a function
of reverse bias voltage when $\alpha$ particles irradiating on the junction
side of the detector, while the lower panel for $\alpha$ particles irradiating
on the ohmic side.}
\end{center}

  As the range of 5 MeV $\alpha$ particles in silicon is about 30 $\mu$m, just a
tenth of the whole thickness of Si bulk, it may not be persuasive to determine
the full depletion of p-n junction simply according to the signals from incidence
on junction side. The lower panel in Fig.~\ref{fig4} presents the same curves
but $\alpha$ particles irradiating on the ohmic side. In this case, under the
bias less than -10 V, the raising time of that signal is rather slow and the pulse
height is slightly small in contrast with the situation of illumination
on junction side. Obviously, it is resulted from the weak electric field near
the ohmic side. The pulse height gets to be saturated and the rise time grows
slowly until the bias increases to -16 V, under which bias voltage, we can
conclude, the p-n junction is fully depleted. Considering that the detector
should be over-depleted, we applied the bias voltage of -20 V across the
detector during the test. Under this bias, the raising time of signal of
$^{241}$Am $\alpha$ particles is 158 ns.

\section{Detection performance}
\subsection{Energy resolution}
  Figure~\ref{fig5} presents a typical $\alpha$ particle energy
spectrum of $^{239}$Pu source. The two peaks are corresponding to 5.157 MeV
and 5.499 MeV~\cite{rytz}, respectively. The peak of 5.499 MeV is caused by
the mixed $^{238}$Pu, and the peak of 5.157 MeV contains the contributions
from 5.144 MeV and 5.106 MeV $\alpha$ particles of small branch ratio but
being unable to be distinguished. After single-peak Gaussian fit of the 5.157 MeV
peak, one can calculate that the $\sigma =10.484$ Ch and the peak is located
at 2545.8 Ch, resulting in the energy resolution of 0.97\%. That is to say
the energy resolving capability is about 50 keV. The energy resolution was
relatively consistent across the four quads with a difference less than 5\%
off the average. Considering that the $\alpha$ source is too close (about 8 cm)
to the detector that the incident angle tends to vary largely so as to increase
the energy straggling, the energy resolution may improve under the condition
of the vertical incidence after collimation. The performance of energy
resolution is acceptable.

  It is the good energy resolution that is the most dramatic factor of silicon
detector. And it ensures the wide applications of these detectors on energy
spectrum measurements and particle discrimination. Energy resolution depends
on many physical factors, not only the detector itself, but also the front-end
electronics system and even the energy and mass of the charged particle to
be detected. If the impacts of system energy resolution are excluded, the
intrinsic energy resolution of the detector should be better than 50 keV.

\begin{center}
\includegraphics[width=7cm]{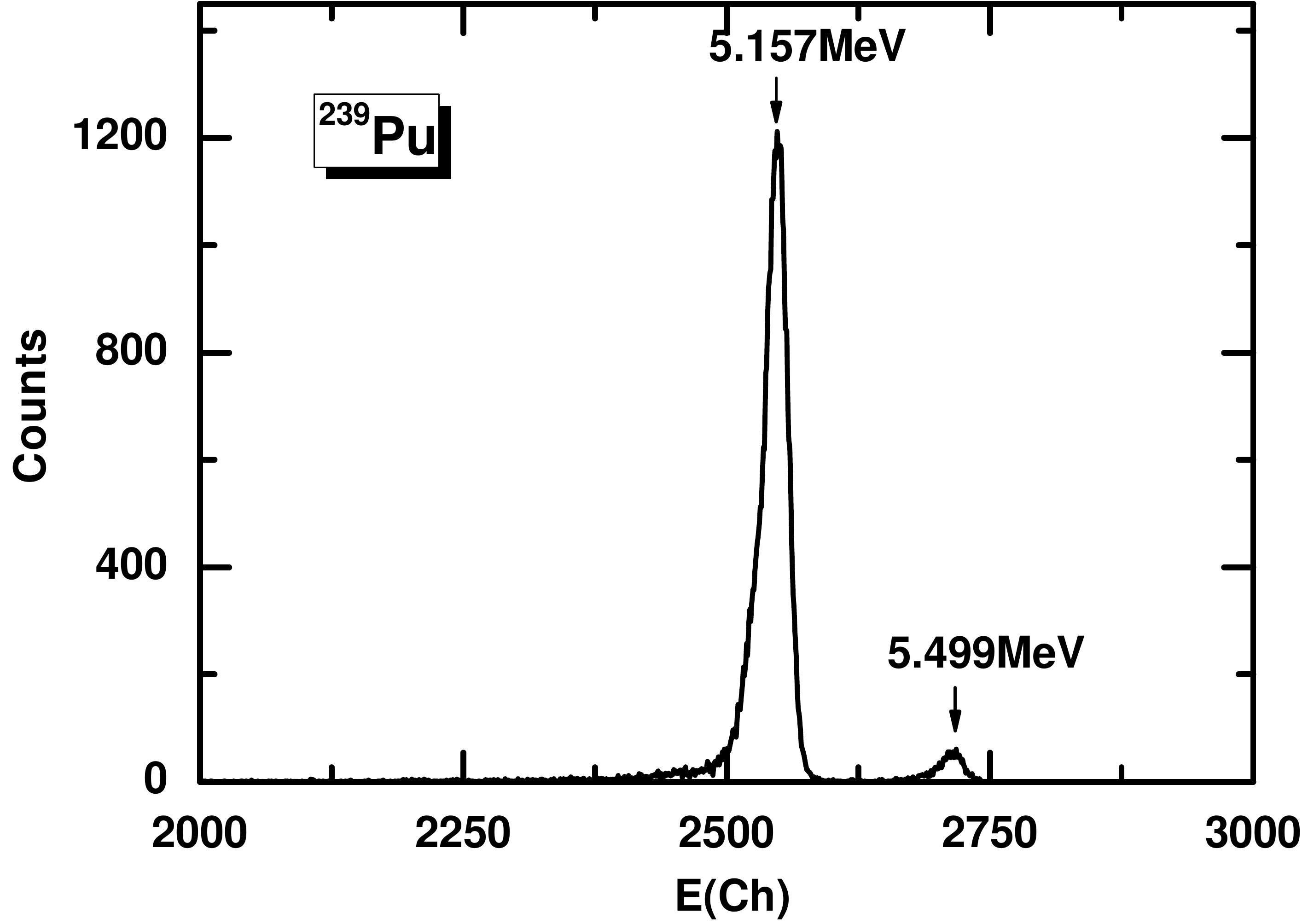}
\figcaption{\label{fig5}Typical $^{239}$Pu $\alpha$ particles energy spectrum of
the QSD. The two peaks are corresponding to 5.157 MeV and 5.499 MeV, respectively,
with energy resolution of 0.97\%.}
\end{center}

\begin{center}
\includegraphics[width=7cm]{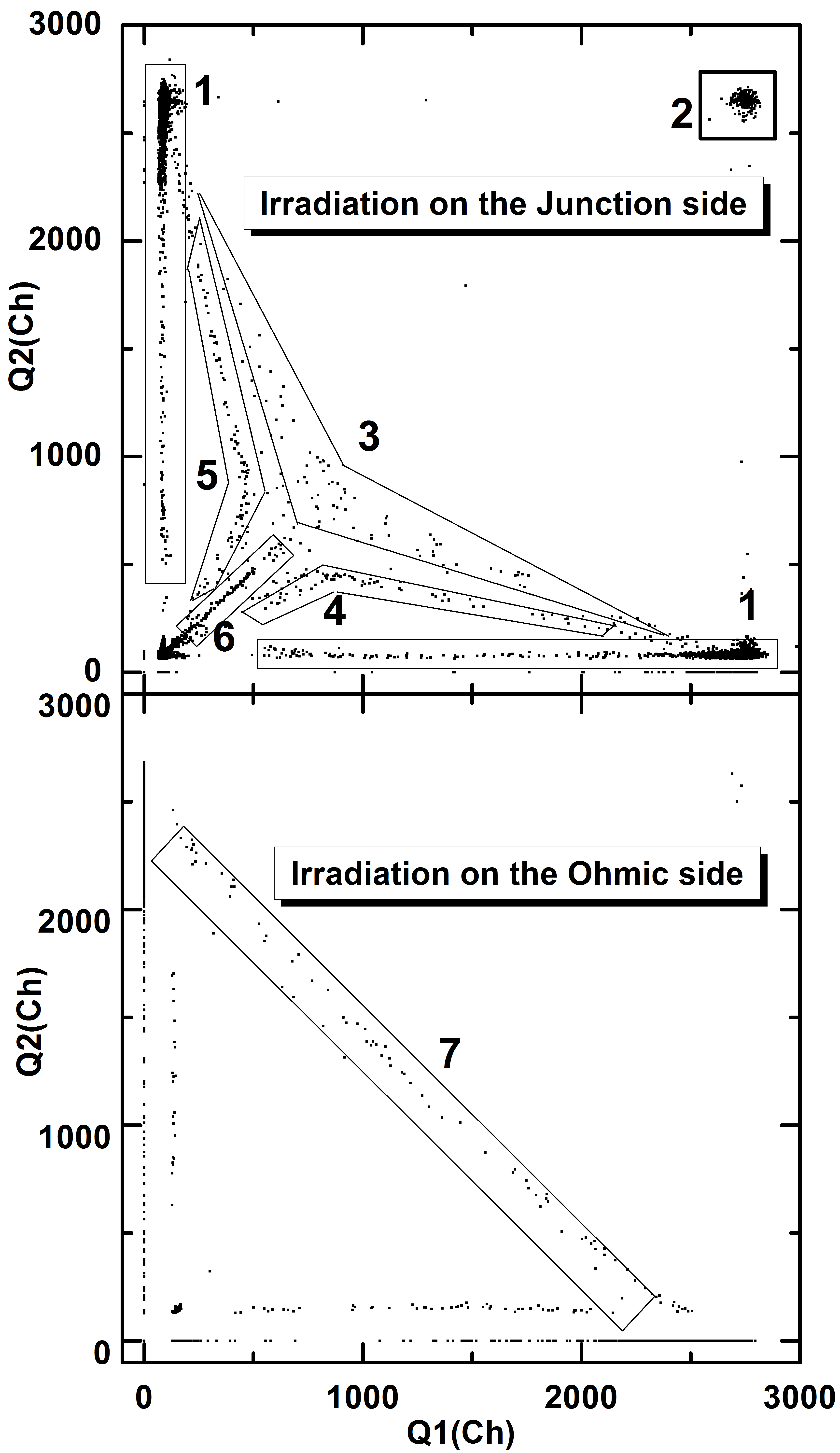}
\figcaption{\label{fig6}Typical correlation spectrum of $^{241}$Am $\alpha$ particles
between two neighbouring quads on the same QSD, the upper panel for $\alpha$ particles
irradiating on the junction side and the lower panel for the ohmic side. See
context for more details about the origination of these correlations.}
\end{center}

\subsection{Charge collection analysis}
  Figure~\ref{fig6} shows typical correlation spectrum of $^{241}$Am $\alpha$
particles between two neighbouring quads on the same QSD, the upper panel
for $\alpha$ particles irradiating on the junction side and the lower panel
for the ohmic side. Charge sharing effect~\cite{poehlsen, yorkston, grassi}
can be observed from both of them. For convenience of expression, the
spectrum are divided into seven zones.

  In the upper panel, the vast majority of events (in Zone 1, almost 98.9\% of
all effective events) shows the two quads records $\alpha$ particles independently
in addition to few accidental coincidence events (in Zone 2, about 0.4\%). There
are also a tiny number of events (in Zone 3, 4, 5, and 6, about 0.6\%) show
complicated correlations between the two quads. The active area of the detector
is 48$\times$48 mm$^2$, and the width of the isolation bar is 0.1 mm, then
one can get the ratio of dead zone to the active zone is $(0.1+0.1)\div 48=0.42\%$.
From its consistence with the ratio of the number of correlation events to
that of total events, one can surmise that the correlation may be caused by
events that $\alpha$ particles incident on the inter-quad area.

  It encourages us to do further test to search out the origination of these
correlation events. As shown in Fig.~\ref{fig7}, different part of the
detector was masked by a thick paper board with a window at different
position to let $\alpha$ particles merely irradiate on specific area,
events from which can be distinguished therefore.

\begin{center}
\includegraphics[bb= 0 0 716 381, width=7cm]{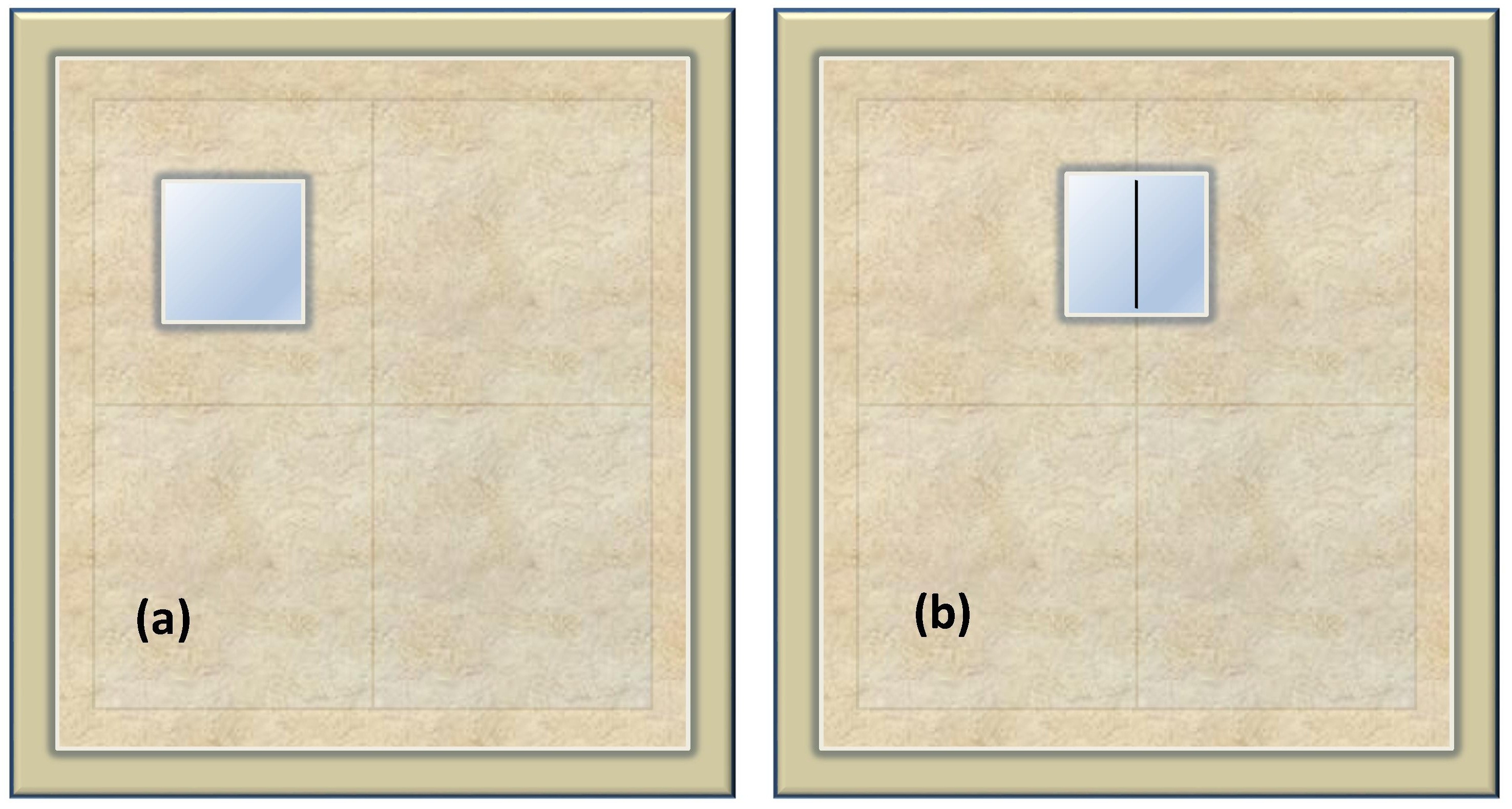}
\figcaption{\label{fig7}Two test modes with a windowed thick paper board masking
different part of the detector. The light blue area can be irradiated by $\alpha$
particles. During these tests, the $\alpha$ source faced the window to
avoid oblique incidence.}
\end{center}

  In Fig.~\ref{fig7}(a), only part of one quad was irradiated. This test shows
that there is no correlation observed between any two quads. The spectrums of
three non-irradiated quads were blank except noises in extremely low channels.
It is concluded that the front-to-end electronic crosstalk can be neglected
and the effects of capacity coupling were not observed.

\begin{center}
\includegraphics[width=7.5cm]{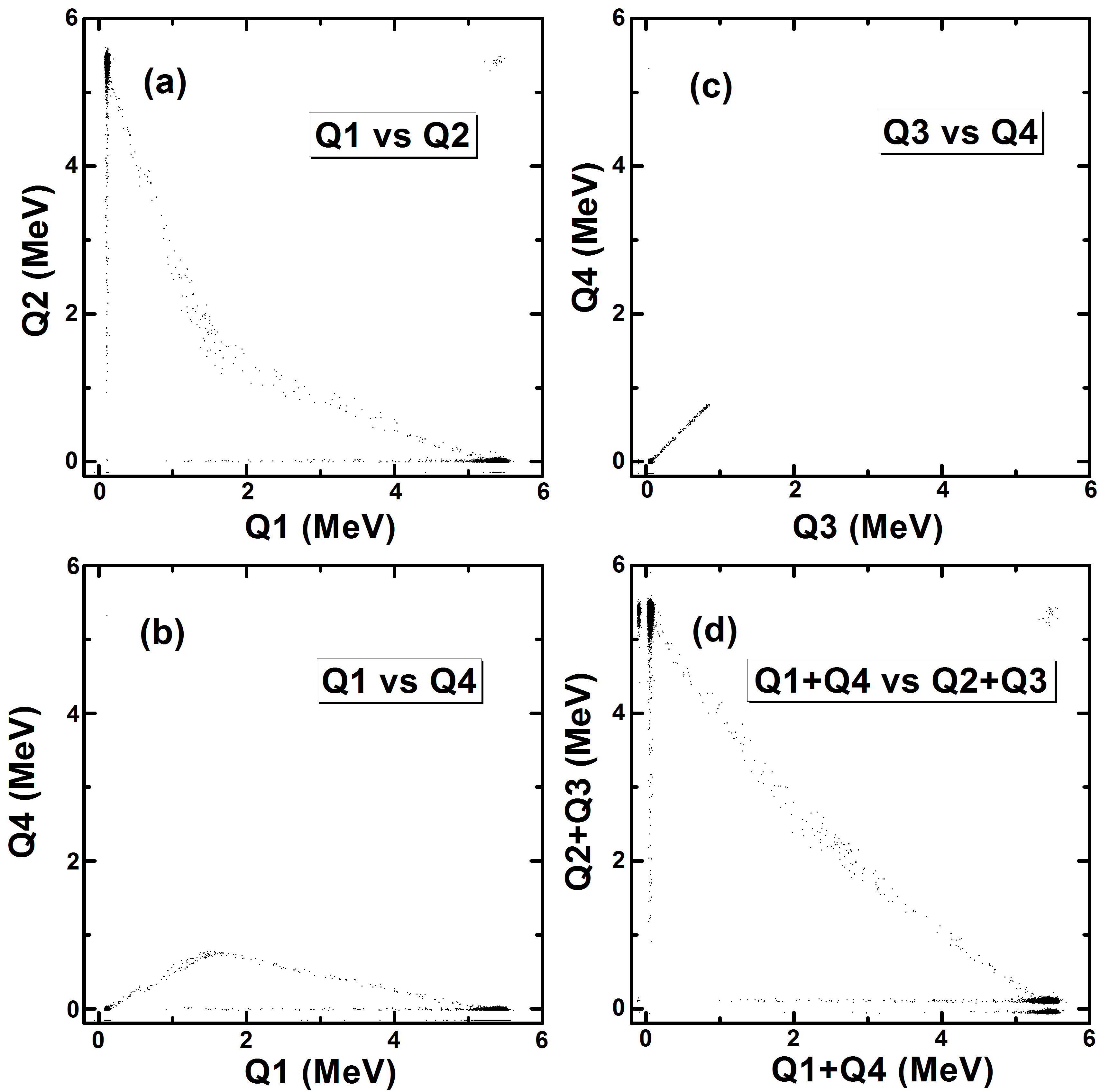}
\figcaption{\label{fig8}Correlation spectrum of test in Fig.~\ref{fig7}(b).}
\end{center}

  As for Fig.~\ref{fig7}(b), part of the inter-quad region between two quads was
irradiated, and events originated from this region can be studied. Four panels
in Fig.~\ref{fig8} present the scatter plots of different pairs of the four
quads. In Fig.~\ref{fig8}(a), symmetrical broken-line of spectrum of Q1 vs Q2
is presented. Obviously, charges ionized by $\alpha$ particles that irradiated
on the inter-quad region are not collected completely by Q1 and Q2, otherwise
the spectrum of the two quads should be a straight line $E_{Q1}+E_{Q2}=E_{tot}$.
Furthermore, the closer to each other the values of Q1 and Q2 become, the greater
quantity of charges they lose. The lost charges are collected by Q3 and Q4, as will
be further discussed below. That is certainly corresponding to the events in
Zone 3 of Fig.~\ref{fig6}.

  From the plot of Q1 vs Q4 in Fig.~\ref{fig8}(b), one can find out that Q4 gets to
its maximum at the turning point of the broken line of Q1 vs Q2, meaning that Q4
collects a maximum amount of charge when Q1 and Q2 losing the most. It can be
concluded that charges are shared by four quads even $\alpha$ particles just
incident on the inter-quad region between two quads. It is similar to the events
in Zone 4 and Zone 5 of Fig.~\ref{fig6}. Zone 4 is composed of events that $\alpha$
particles irradiated on the inter-quad area between Q1 and Q4, and the charges are
partly collected by Q2. Likewise, events in Zone 5 stand for $\alpha$ particles
irradiating between Q2 and Q3, and part of the charges are collected by Q1.

  The plot of Q3 vs Q4 is given out in Fig.~\ref{fig8}(c). What puzzled a lot is
that Q3 always collected the same quantity of charges with Q4. It may be
understood like that the charges ionized at the inter-quad region between
Q1 and Q2 may not only be collected by Q1 and Q2, but also drift along the
long inter-quad insolation strip and finally be bisected by Q3 and Q4.
Homologous events in Fig.~\ref{fig6} are located in Zone 6, that charges are
ionized between Q3 and Q4, and part of them drifting along the SiO$_2$ isolation
strip are equally shared by Q1 and Q2 finally.

  In order to further verify this, we give out the spectrum of Q1+Q4 vs Q2+Q3 in
Fig.~\ref{fig8}(d), the correlation between these two parameters are closer
to the straight line. According to this fact, it can be concluded that most of
the charges are shared by the quads on the both sides of the isolation strip,
and only a small part of charge were lost during the collection.

  Unlike the case in upper panel of Fig.~\ref{fig6}, events of $\alpha$ particles
irradiating on the ohmic side do not have complex correlations. The straight
line in Zone 7 represents events that charges are completely collected by two
neighbouring quads, but not shared by four quads.

  Conclusions can be made based on different phenomenons of irradiation on two
sides of the detector, that the electric field distortion occurs in the inter-quad
regions, and only near the surface of junction side. $\alpha$ particles irradiate
on the inter-quad region between two quads on junction side and the ionized
charges are affected by the distorted electric field and partly collected by
the other two quads. While irradiation take place on the ohmic side, $\alpha$
particles cannot feel the electric field distortion before being stopped, and the
ionized charges were completely collected.

  Generally speaking, the amount of the inter-quad events are small enough to
be neglected in practical use of the experiments.

\section{Summary}
  The QSDs for charged particle detection were developed at China Institute of
Atomic Energy. The thickness of the QSD is about 300 $\mu$m and the active area is
48$\times$48 mm$^{2}$ with 0.1 mm wide isolation bar between each two quads.
Tests with $^{241}$Am and $^{239}$Pu radioactive sources were carried out to
assess the performance of the detector. The leakage current under the over-depletion
bias voltage of -20 V is as small as 5.01 nA, and the raising time is better than
160 ns. The energy resolution for 5.157 MeV $\alpha$-particle is about 1\%. The
analysis results show the existence of complicated correlations between neighboring
quads, which was caused by the electric field distortion near the isolation bar.
The inter-quad events would not impact too much as the amount of these events is tiny.
These performances are acceptable. In view of the fabrication reaching
such a level that these detectors can be used in nuclear physics experiments,
one should also consider the long term behaviour of these detectors, such as
radiation damage and electrical stability.
\end{multicols}

\vspace{10mm}

\vspace{-1mm}
\centerline{\rule{80mm}{0.1pt}}
\vspace{2mm}

\begin{multicols}{2}

\end{multicols}

\clearpage

\end{document}